\documentclass[acmsmall,screen]{acmart}

\begin{document}

\title{Characterizing disruptions in online gaming behavior following software patches}

\author{Xiaozhe `Arcadia' Zhang}
\affiliation{%
  \institution{Department of Computer Science, University of Colorado Boulder}
  \country{United States}}
\email{arcadia.zhang@colorado.edu}

\author{Brian C. Keegan}
\affiliation{%
  \institution{Department of Information Science, University of Colorado Boulder}
  \country{United States}}
\email{brian.keegan@colorado.edu}

\renewcommand{\shortauthors}{Zhang and Keegan}

\begin{abstract}
Multiplayer online games are ideal settings for studying the effects of technological disruptions on social behavior. Software patches to online games cause significant changes to the game's rules and require players to develop new strategies to cope with these disruptions. We surveyed players, analyzed the content of software patch notes, and analyzed changes to the character selection behaviors in more than 53 million matches of \textit{Dota 2} in the days before and after software patches over a 30-month period. We found that the severity of patches is correlated with the magnitude of behavioral changes following a patch. We discuss the opportunities of leveraging software patches to online games as a valuable but overlooked empirical instrument for measuring behavioral dynamics.
\end{abstract}


\keywords{online games, patch severity index, Defense of the Ancients 2, Dota 2}

\maketitle

\section*{Introduction}
How do software patches impact the behavior of users? Online games are highly interoperable systems that require on-going maintenance for security, usability, balance, and new features~\citep{chapin_types_2001}. The release of a software patch is necessarily disruptive as hardware systems need to be rebooted, software dependencies need to be refreshed, user routines need to be revised, and documentation needs to be updated. In most software systems, users pay little attention to patched beyond interacting with notificationss~\citep{fagan_study_2015,vaniea_tales_2016,bergman_cognitive_2018}: they generally do not read change logs, index their experience around version numbers, or create or engage with social media content. However, the experience of software patches in online game communities is significantly different. Players index their experience using patch notes, actively lobby for and organize against changes, and produce social content both in anticipation of and in reaction to software patches. 

Multiplayer online battle arenas (MOBAs) like \textit{League of Legends} (\textit{LoL}) and \textit{Defense of the Ancients 2} (\textit{Dota 2}) are ideal settings to examine the effects of software patches on user behavior for several reasons~\citep{bainbridge_creative_2007,gursoy_understanding_2019}. First, MOBA players consciously change their behavior in response to patches in order to remain competitive: this gives us a credible and exogenous mechanism to understand the causal relationship between patches and behavior. Second, MOBA games contain public, detailed, and longitudinal records of within-subject behavior: this enables us to compare player behavior before and after patches. Third, historical records about the timing and content of the patches are well-documented: this variation in patches could explain variation in players' in-game behavior. Finally, players are often very active in online communities participating in online discussions, so it is feasible to recruit for surveys or interviews. However, the relationship between software patches as a site of technological change and the effects of these changes on post-patch behavior in a globalized gaming culture remain poorly understood. 

This papers explores how players' character selection behavior is temporarily disrupted following a patch and how the magnitude of this behavioral change varies with the ``severity'' of the patch's changes. First, we surveyed \textit{Dota 2} players to understand how they change their in-game behaviors following software patches. Second, we developed a ``patch severity index'' using a combination of survey data and patch note content to quantify the magnitude of a patch's changes. Third, we used a public dataset of \textit{Dota 2} players' character selections over a 36-month period of time to measure changes in character selection behavior around patches. Our survey results confirm that players in-game behaviors are strongly influenced by software patches, there is significant variation in the ``severity'' of the changes introduced by software patches, there are significant but temporary disruptions in players' character selection behavior around the patches, and the scale of these disruptions are strongly correlated with severity of patches.

\section*{Background}
\subsection*{Multiplayer online battle arenas}
Team-based multiplayer online games offer many benefits for research about social behavior, making them potentially better empirical settings into team processes outside of laboratories~\citep{elnasr_game_2013,williams_mapping_2010,yee_labor_2006}. Multiplayer online battle areas (MOBAs) like \textit{League of Legend} and \textit{Dota 2} are one of the most popular genres of multiplayer online games. In the canonical game mode, ten players of similar skill levels will be paired together into two teams of five players. Over the course of an approximately 20--45 minute match, each player controls a single character\footnote{We use the terms ``heroes'' and ``characters'' interchangeably for the in-game agents that players select and control.} that accumulates experience and gold from destroying enemy buildings, characters, and non-player characters. The experience is redeemed by leveling up new abilities and gold is redeemed by purchasing items augmenting or complementing their characters' abilities. The match ends when one team's base is destroyed or one team resigns. Before the game starts, players go through a drafting phase where they pick and ban characters out of more than 100 available choices. Drafting plays a significant role in MOBA as it determines every player's roles the team's overall strategy, and is critical for the team's success. In a MOBA setting like \textit{Dota 2}, characters vary significantly in their abilities, which allows them to fulfill different roles such as damage dealers (``carries''), damage takers (``tanks''), and specialists (``supports''). 

MOBAs like \textit{Dota 2} exhibit high levels of decision complexity for players: differentiated character roles requires strong coordination, discretion in executing in-game tasks requires trust, and the mutual observability of team designs requires strategic meta-cognition. Previous work has looked at team composition strategies~\citep{sapienza_deep_2018,guo_analysis_2012,kim_strong_2017}, team behavior pattern mining~\citep{drachen_skill-based_2016,li_visual_2017,yang_identifying_nodate}, and the impact of the metagame~\citep{peabody_detecting_2018,donaldson_mechanics_2017,debus_metagames_2017}. Role specialization as well as other pre- and in-game tactics like situational awareness and meta-cognition about opponents are referred to as a ``meta-game''~\citep{kim_proficiency_2016,donaldson_mechanics_2017,debus_metagames_2017}. Changes like adding new characters, changing existing characters' abilities, or altering the map cause change incentives, tactics, and other relationships throughout the game.

\subsection*{Software patches}
Patches have always been a tool for software developers to fix security vulnerabilities, provide improvement and add new features to existing software infrastructures~\citep{dadzie_understanding_2005}. Video game development differs from other software development in many ways~\citep{pascarella_video_2018,saiqa_aleem_game_2016,murphy-hill_cowboys_2014,ampatzoglou_software_2010,lewis_repairing_2011,lin_studying_2017}. The developers of massively multiplayer online games (MMOGs) and multiplayer online battle arenas (MOBAs) engage in an on-going process of ``balancing'' to ensure characters do not become over- or under-powered and reduce the fairness of competitiveness of matches~\citep{daneva_striving_2017,lewis_what_2010}. Software patches increase (``buff'') or decrease (``nerf'') the power of existing characters' and items' abilities as the game developers strive to achieve a ``balance''. Patches introduce new characters and items, and alter other features such as the map, non-player characters, or experience and economy mechanics~\citep{claypool_impact_2017,gursoy_understanding_2019,kica_nerfs_2016}. 

While the official patch notes are dry technical communication, they are still actively framed by developers~\citep{mason_video_2013,sherlock_patching_2014}. The power to require users to install developer updates to access a community is a form of ``protocological power'' making games highly contingent commodities~\citep{svelch_resisting_2019}. Game developers are required to engage in an on-going process of data collection and monitoring in order to use their power to unilaterally introduce changes to an experience millions of people are invested in to responsibly govern community values like fairness~\citep{malaby_coding_2006,chen_changing_2008}. Software developers can face the resistance from their user communities when releasing patches, which can sometimes lead to coordinated campaigns on social media platforms~\citep{acker_software_2016} and negative reviews on app stores~\citep{franzmann_influence_2019}. 

In addition to their technical component, software patches in online games are also settings for social behavior like anticipation, deliberation, and sensemaking. Before a patch is released, players lobby developers for changes through social media, speculate about the content and timing of the release, and even write-up fake patch notes for humorous or trolling purposes. Players create hours-long YouTube videos interpreting the patch changes line-by-line, players reference different epochs of in-game experience using patch numbers, and platforms like Reddit, Twitter, and Twitch are flooded with commentary about the patch~\citep{burger-helmchen_user_2011}. This is perhaps captured most vividly by players of multiplayer online game indexing their experience playing the game over time through release numbers (\textit{e.g.}, ``Back in 6.89...''). Because software patches introduce many changes to the game, they disrupt players' knowledge, strategies, and behaviors and require them to adapt and improvise new alternatives following a patch. These problems are non-trivial to solve because of the multiple dimensions and complex interactions among characters' abilities and roles, game mechanics, and players' meta-strategies~\citep{debus_metagames_2017,donaldson_mechanics_2017}.

\subsection*{Hypotheses}
Players' behavior in the aftermath of a patch should capture the traces of their problem solving processes as they discover, adapt, and emulate strategies. 
Prior research has analyzed changes in character usage and banning behavior following patches~\citep{peabody_detecting_2018,zhang_investigating_2017,he_effects_2021,wang_research_2021}. This research used \textit{League of Legends} as an empirical setting, a game that employs a ``freemium'' business model where players have to purchase characters in order to play them. In contrast, all characters are free-to-play for players in \textit{Dota 2}, which ensures that players can freely select any character following a patch. Using \textit{Dota 2} as empirical setting removes an important confounder in explaining how players' character selection tactics are affected by software patches: the availability of \textit{all} characters to \textit{all} players before and after a patch in \textit{Dota 2} enables us to measure the effects of a patch on \textit{both} the patched characters (where post-patch changes in behavior are expected) as well as in un-patched characters (where post-patch changes in behavior are emergent). 

Because of the difficulties of understanding the space of all solutions and the strong incentive to remain competitive against other players, there are many lines of inquiry to explore with patched-induced behavioral changes. We focus on changes in character selection behavior because survey respondents (see Table~\ref{tab:follow_patches}) indicated this was one of the most common behavioral changes. We make two hypotheses about the main effects of players' character selection behavior in the aftermath of software patches. First, players are likely to employ strategies that produce ``good enough'' strategies: we expect to see the diversity of players' character selection behaviors decrease significantly in the aftermath of a path as they resort to risk-averse strategies. Second, patches with more changes will introduce more uncertainty into this complex information space than a patch with fewer changes. We expect that the magnitude of the behavioral changes will vary with the disruptiveness of a patch: more ``severe'' patches should cause greater post-patch behavioral changes than less ``severe'' patches.

\section*{Methods}
\subsection*{Log data description}
We collected our in-game match data from two sources: (1) a data dump released by Opendota.com which contains 1,191,768,403 matches dated up to March 2016~\cite{cui_dump_2016} and (2) log data from March 2016 to July 2016 from \textit{Dota 2}'s official API~\cite{dota2_api}. We only selected matches played in games modes where there is no system randomization on character picking strategy: ``All pick'', ``Captain's Mode'', ``Ranked Matchmaking'', and ``Compendium Matchmaking''. Examples of game modes we excluded include ``All Random'', in which each player gets assigned a random hero, and ``Single Draft'', in which each player can only choose from three random characters. We filtered the data to analyze all matches played 30 days before and after each major patch. The final dataset contains 532,057,466 matches, spanning from patch 6.78 (May 2013) to patch 6.88 (June 2016). While player skill level is reported in the data after May 2015, it was not capture beforehand, thus omitting a 7 of the 11 patch events in our time frame. Because reconstructing and validating players' historical skill levels around patches would computationally expensive and difficult to validate, we do not stratify our research design by skill level and only report aggregate results across all skills. Other meta-data includes the unique match ID, the game mode, the start time of the match, the ten players' IDs, each player's team assignment, and each player's character selection.

\subsection*{Survey}
In addition to the in-game match data, we deployed a survey to the \textit{Dota 2} online community inquiring about players' patch-related behavior. Following approval by our institutional review board as well as community moderators, the survey was advertised through Reddit, Twitter and Discord servers. After soliciting responses for two weeks, incomplete and spam responses were filtered out, and we had a total of 449 valid responses. Our sample shows an extreme gender bias mirroring general demographics in MOBAs~\citep{ratan_league_2012,ratan_stand_2015}: 96.4\% of the survey participants identified as male, 1.6\% identified as female and 0.9\% identified as gender diverse. Most participants were below 35 years old: 59.6\% participants indicated that they are 18 to 24 years old and 36.3\% are 25 to 34 years old. 88.8\% of participants have spent more than 1000 hours in the game \textit{Dota 2} and 66.2\% have been playing the game since 2014. In addition to demographic questions, the survey asked respondents about their history and experience of playing \textit{Dota 2}, their information seeking behavior following a patch, the importance they assign to different kinds of changes in patches, and their character picking and play strategies before and after patches. We report out the survey results in the context of other findings in the Results section below.

\subsection*{Character selection behavior}
Patches introduce changes along multiple dimensions of an already complex and multidimensional decision space. Character selection is a crucial step in the experience of playing the game and poor team assembly decisions can severely reduce performance, cohesion, and likelihood of winning~\cite{kou_playing_2014,kim_proficiency_2016,summerville_draft_2016,semenov_performance_2016}. Because the vast majority of players (89.5\%) in our survey reported changing their character selection behaviors following patches (Table \ref{tab:follow_patches}), we prioritized character selection as the primary behavior to analyze for disruptions around software patches. There are many other kinds of patch-induced behavioral changes that could be detected using the match data (items, character combinations, game length, kills/deaths/assists, \textit{etc}.) and there are good reasons to suspect these changes vary across social categories (skill, geography, \textit{etc}). Given the limitations of the data described above for segmenting the users into these categories, we were not able to explore these additional features and only report out results aggregated across the entire population of matches in this time frame. We describe opportunities for exploring other patch-induced behavioral changes in the Discussion section.

\begin{table}[t]
\caption{Two questions asking participants about whether or not they follow patch notes and whether patch notes influence their character picks} 
\footnotesize
\centering
\begin{tabular}{lcc}
\toprule
Question & \% Yes & \% No \\ \hline
Think about a day when a new patch is released, do you look for information about the patch? & 98.9\% & 1.1\% \\
Do you adjust your hero picks based on what you learn from patch notes or playing the game? & 89.5\% & 10.5\% \\
\bottomrule
\end{tabular}
\label{tab:follow_patches}
\end{table}

To measure these patch-induced behavioral changes, we condensed character selection behavior into daily aggregated frequency vectors for 30 days before and after each major patch. We chose this 30-day window before and after a patch to establish a baseline for character selection behaviors and then the measure whether and how long it takes for character selection behavior to stabilize following the patch as players converge on new meta-strategies. In other words, we expect that any behavioral changes caused by the patch should have stabilized by 30 days after the patch. For each major patch from 6.78 to 6.88, we constructed a matrix $M$ of days and characters: $M_{ij}$ is the number of times character $j$ is picked on the $i^{th}$ day, where $i \in [-30, 30]$ around the patch release and $j \in [1, 102]$ for the number of unique characters in the game. For new characters that were added later into the game through patches, they have a pick count of 0 for the patches that they were not present.

\subsubsection*{Cosine similarity.} 
\begin{figure}[t]
\centering
\caption{A schematic diagram illustrating the changes in player behavior before the patch ($\Delta t - 1$) and immediately following a patch ($\Delta t + 1$) compared to 30 days following a patch.}
\label{fig:delta_example}
\includegraphics[width=.75\textwidth]{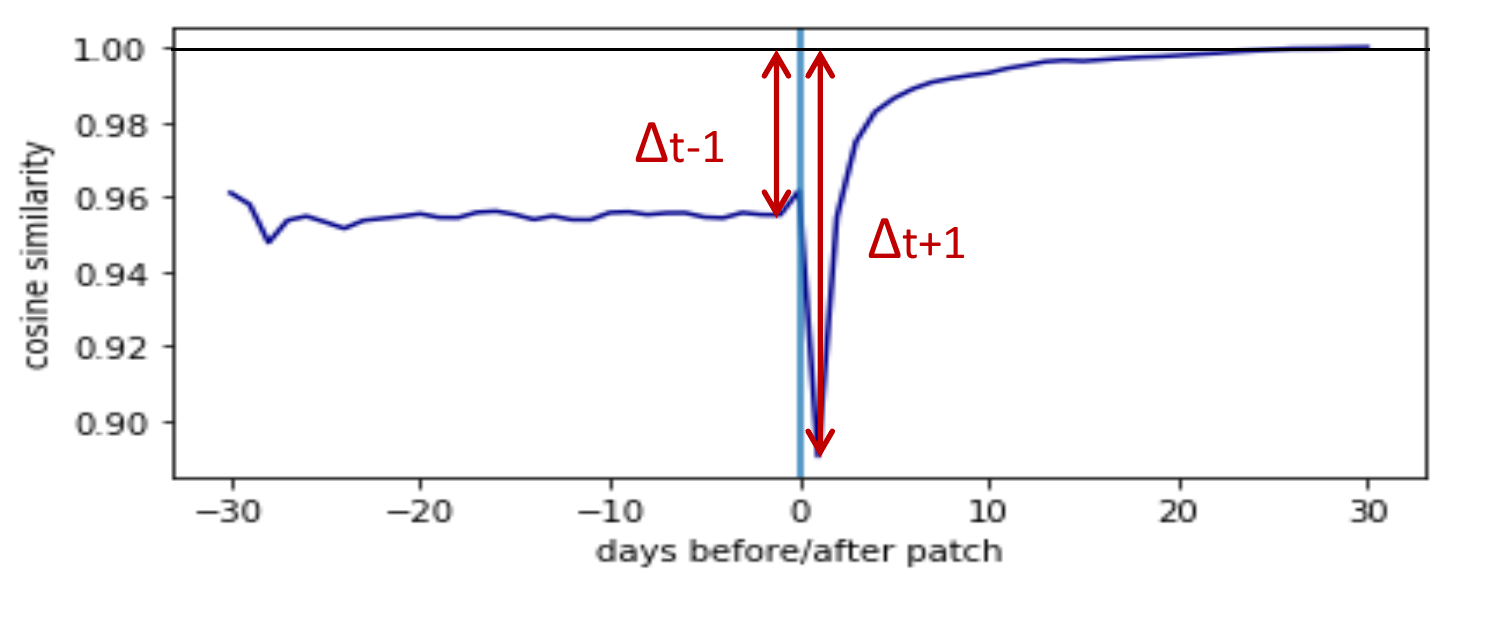}
\end{figure}

Cosine similarity is used to measure similarities between two non-zero vectors. We used cosine similarity to determine the similarity of the aggregated daily draft vectors compared to the draft vector 30 days after a patch. The $\Delta^{cosine}_{t-1}$ and $\Delta^{cosine}_{t+1}$ values are the differences in the cosine similarities of player selections the day immediately before and after (respectively) the patch compared to 30 days after the patch (see Figure~\ref{fig:delta_example}). $\Delta^{cosine}_{t-1}$ captures the change in character selection behavior from the day immediately prior to the patch to 30 days after the patch. $\Delta^{cosine}_{t+1}$ captures the change in character selection behavior immediately following a patch. Low similarity scores mean that players' character selection behavior was substantially different than 30 days after the patch. We constructed character pick matrices for each of the 11 patches (6.78--6.88) and calculated the similarity between each day's draft for the 30 days before and after each patch. Note that normalization of character selection is not necessary because the normalization terms will cancel out per the formula of cosine similarity.

\subsubsection*{Gini coefficient.} Gini coefficients the equality of player attention across characters. Gini coefficients range from 0 to 1, where 0 represents perfect equality (every character is picked with the same frequency), and 1 represents perfect inequality (every player picks the same hero, which is impossible in \textit{Dota 2} since players cannot pick the same character in a match). We constructed character pick matrix for the 11 major patches (6.78 -- 6.88), and we were able to generate 11 vectors of 61 values representing the Gini coefficients of character picks for the 30 days before and after the patch. The $gini_{t-1}$ value is the Gini coefficient of character selections the day immediately before a patch and $gini_{t+1}$ is the coefficient of character selections the day immediately following a patch. For example, a decrease in cosine similarity and an increase in the Gini coefficient implies that even though players' strategies are becoming more diverse, these strategies still prioritize selecting from a small set characters rather than across many characters. A decrease in the Gini coefficient implies that players' drafts are spread more evenly across the set of possible characters.

\subsection*{Patch severity index}

A patch note is a document released by the game developers enumerating the specific changes made since the previous version~\citep{bainbridge_creative_2007,claypool_impact_2017,gursoy_understanding_2019,kica_nerfs_2016,mason_video_2013,sherlock_patching_2014}. In MOBAs like \textit{Dota 2} and \textit{LoL}, these changes can occur at multiple levels involving characters, items, maps, gameplay, and other features~\citep{gursoy_understanding_2019}. For example, changing the location of a tree on the map does not have the same impact on a player's character selection behavior as changing the damage dealt by a hero. We define a ``change'' to be a single instance of any one of these documented differences in developers patch notes. Because some changes have a greater impact on gameplay than others, changes cannot be treated equally. Some examples of patch changes are given in Table~\ref{tab:patch_change_examples}.

\begin{table}[t]
\footnotesize
\centering
\caption{Examples of patch changes by category.}
\begin{tabular}{lcl}
\toprule
\textbf{Category} & \textbf{Patch} & \textbf{Example change} \\ \hline
  \textit{Hero} & 6.86 & ``Enchantress: Enchant cooldown reduced from 30/25/20/15 to 30/24/18/12'' \\
  \textit{Item} & 6.86 & ``Quelling Blade: Cost reduced from 225 to 200'' \\
  \textit{Economy} & 6.84 & ``AoE Bonus Gold component based on Team Net Worth difference reduced by 25\%'' \\ 
  \textit{Experience} & 6.84 & ``AoE Bonus XP component based on Team XP difference reduced by 40\%'' \\
  \textit{Map} & 6.86 & ``Moved the Radiant hard camp closer to the Dire offlane'' \\
  \textit{Neutral} & 6.86  &  ``Centaur Courser now has a stacking magic resistance aura'' \\
  \textit{Other} & 6.86 & ``Added Arcane Rune'' \\
  \bottomrule
\end{tabular}
\label{tab:patch_change_examples}
\end{table}

\begin{table}[t]
\caption{Features of the Patch Severity Index.}
\footnotesize
\centering
\begin{tabular}{lllc}
  \toprule
  \textbf{Category} & \textbf{Description} & \textbf{Weight}  \\ \hline
  \textit{Hero} & Changes to heroes abilities or attack damage. & 25/100 \\
  \textit{Item} & Changes to item recipes or costs. & 15/100 \\
  \textit{Economy} & Changes to gold gained after killing heroes. & 10/100 \\
  \textit{Experience} & XP gained per creep and XP needed to level up. & 10/100 \\
  \textit{Map} & Changes made to terrain (Roshan, trees). & 10/100 \\
  \textit{Neutral} & Changes made to jungle neutrals, lane creeps. & 5/100 \\
  \textit{Other} & Other miscellaneous changes (formulas).  & 10/100 \\ 
  \bottomrule
\end{tabular}
\label{tab:psi_features}
\end{table}

We propose a ``Patch Severity Index'' (PSI) a metric to quantify the magnitude of changes introduced by a patch. We collected patch notes from \texttt{liquipedia.net}, a collaborative wiki documenting patch changes among other game information. Patch notes on \texttt{liquipedia.net} originate from the official patch notes released by the \textit{Dota 2} developer Valve but are converted to a consistent formatting across patches by the community. We did a content analysis on the changes across the 11 patches in our sample (6.78 to 6.88) and came up with seven categories summarized in Table~\ref{tab:psi_features}. These categories reproduce similar themes founds in content analyses of patch note documents for \textit{League of Legends}~\citep{gursoy_understanding_2019}. A research assistant categorized all the changes in the patch notes into seven categories based on the content of the changes. Since most of the patch notes are already labeled, either on the official release, or the crowd-sourced \texttt{liquipedia.net}, all categories used in the patch notes are consolidated into the seven categories in Table~\ref{tab:psi_features}. We also counted the raw number of bulleted changes in each patch as a coarse baseline of patch severity. However, for character and item changes, we only count them once for each character and item regardless the raw number of bulleted changes listed.

\begin{figure}[t]
\centering
\caption{Box plots illustrating the distribution of survey responses assigning points to the severity of seven categories of patch changes. The median values used for weights in the Patch Severity Index are labelled in white.}
\includegraphics[width=.75\textwidth]{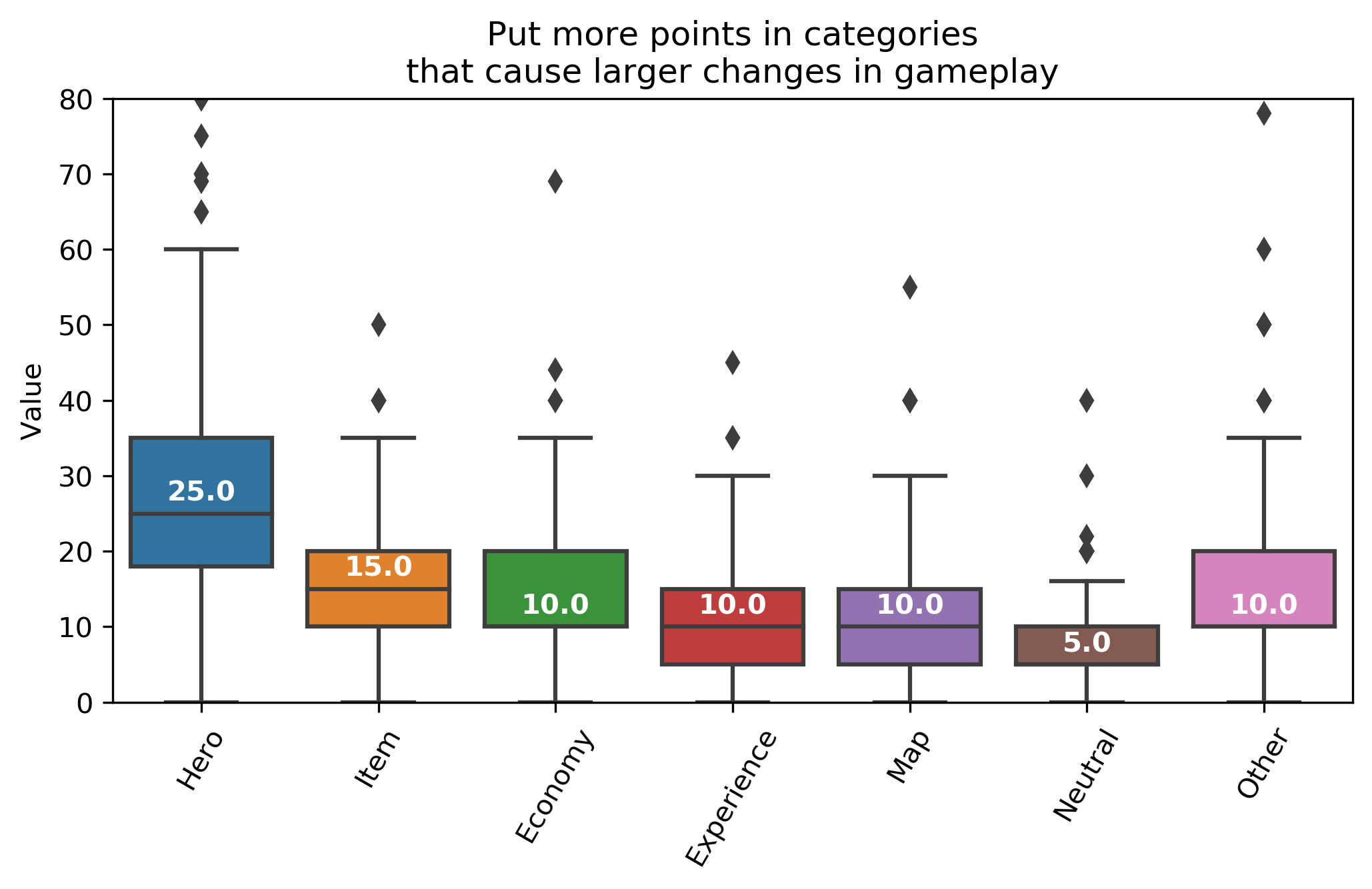}
\label{fig:severity}
\end{figure}

We constructed the PSI as the weighted sum of the changes introduced across categories. We determined these weights from the survey respondents by asking them to distribute 100 points among the seven categories based on their influence on gameplay (``Put more points in categories that cause larger changes in your gameplay''). Figure~\ref{fig:severity} summarizes the results we received in the survey responses. Because of skews in the distribution of respondents' values, we used the median reported values to determine the weights. The final PSI for each patch is calculated as the sum of total number of changes in each category weighted by the median survey response for that category. We normalized the final PSI for each patch against the patch with the highest PSI and the result is shown in Table~\ref{tab:patch_features}.

\begin{table*}[t]
\small
\caption{Summary of the features across the 11 patches in the sample.}
\centering
\begin{tabular}{l|c|cc|cccc|c}
\toprule
  \textbf{Patch} & \textbf{Date} &\textbf{Changes} & \textbf{PSI} & \textbf{$\Delta^{cosine}_{t-1}$} & \textbf{$\Delta^{cosine}_{t+1}$} & \textbf{${gini}_{t-1}$} & \textbf{${gini}_{t+1}$}& \textbf{Match count}\\ \hline
  \textit{6.78} & 2013-05-30 & 121 & 0.69 & 0.010 & 0.007 & 0.274 & 0.269 & 21,552,410\\
  \textit{6.79} & 2013-10-21 & 158 & 0.84 & 0.045 & 0.044 & 0.299 & 0.316 & 34,031,830\\
  \textit{6.80} & 2014-01-27 & 101 & 0.59 & 0.020 & 0.187 & 0.318 & 0.386 & 46,796,895\\
  \textit{6.81} & 2014-04-29 & 94 & 0.59 & 0.012 & 0.014 & 0.331 & 0.333 & 52,001,132\\
  \textit{6.82} & 2014-09-24 & 153 & 0.81 & 0.019 & 0.076 & 0.324 & 0.357 & 55,335,763\\
  \textit{6.83} & 2014-12-17 & 102 & 0.60 & 0.023 & 0.020 & 0.355 & 0.354 & 60,055,781\\
  \textit{6.84} & 2015-04-30 & 190 & 0.96 & 0.057 & 0.040 & 0.410 & 0.383 & 56,441,803\\
  \textit{6.85} & 2015-09-24 & 134 & 0.80 & 0.055 & 0.028 & 0.367 & 0.371 & 50,524,583\\
  \textit{6.86} & 2015-12-16 & 212 & 1.00 & 0.045 & 0.110 & 0.393 & 0.440 & 564,732,84\\
  \textit{6.87} & 2016-04-25 & 189 & 0.90 & 0.185 & 0.180 & 0.400 & 0.415 & 53,316,353\\
  \textit{6.88} & 2016-06-12 & 82 & 0.54 & 0.011 & 0.009 & 0.416 & 0.413 & 58,027,742 \\
  \bottomrule
\end{tabular}
\label{tab:patch_features}
\end{table*}

\section*{Results}

\subsection*{Case study}
We provide a case study of two patches in \textit{Dota 2} to ground the subsequent analyses and discussion. Patches 6.86 and 6.88 are the patches with the highest and the lowest (respectively) patch severity indices out of all patches we studied: patch 6.86 implemented a total of 212 changes while 6.88 introduced 82 changes. Patch changes are broken down by category in Table~\ref{tab:psi_features}: patch 6.88 introduced only two types of changes (item changes and character changes) while patch 6.86 introduced all categories of changes except experience changes. Table~\ref{tab:patch_comparison} provides a detailed comparison of the number of changes introduced by these two patches. 

\begin{table*}[t]
\caption{Number of changes in each category for patch 6.86 and 6.88.}
\small
\centering
\def\tabularxcolumn#1{m{#1}}
\begin{tabular}{cccccccc}
  \toprule
  \textbf{Patch}  &  \textbf{Hero} & \textbf{Item} & \textbf{Economy} & \textbf{Experience} & \textbf{Map} & \textbf{Neutral} & \textbf{Other} \\ \hline
   6.86 & 89 & 53 & 1 & 0 & 47 & 11 & 11 \\
   6.88 & 75 & 7 & 0 & 0 & 0 & 0 & 0 \\ 
   \bottomrule
\end{tabular}
\label{tab:patch_comparison}
\end{table*}

\begin{figure*}[t]
\centering
\caption{Bar plots sorted by characters with the greatest increases (left subplot) and decreases (right subplot) in pick frequency before (blue) and after (orange) patch 6.86.}
\includegraphics[width=\textwidth]{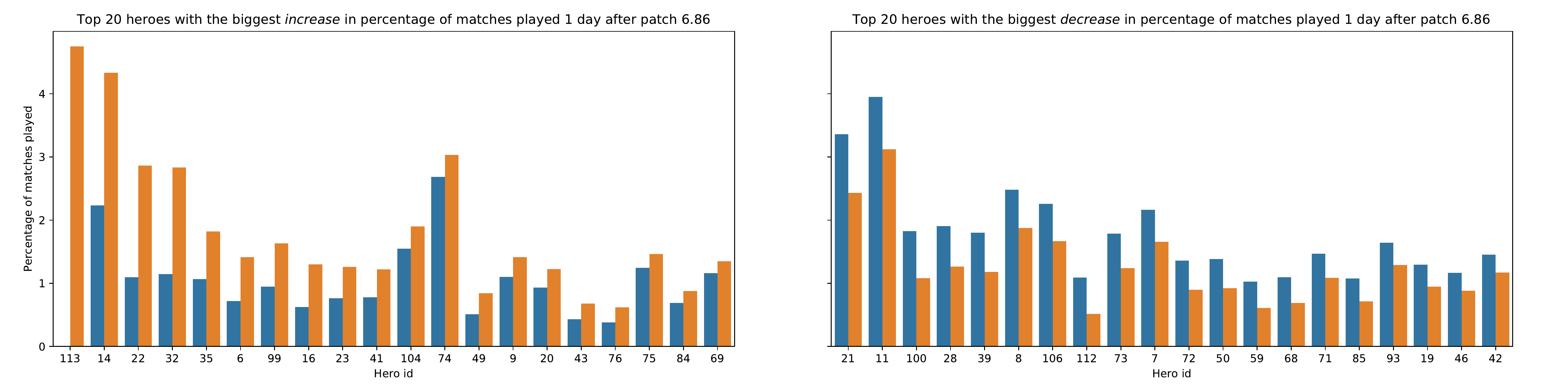}
\label{fig:changes686}
\end{figure*}

\begin{figure*}[t]
\centering
\caption{Bar plots sorted by characters with the greatest increases (left subplot) and decreases (right subplot) in pick frequency before (blue) and after (orange) patch 6.88.}
\includegraphics[width=\textwidth]{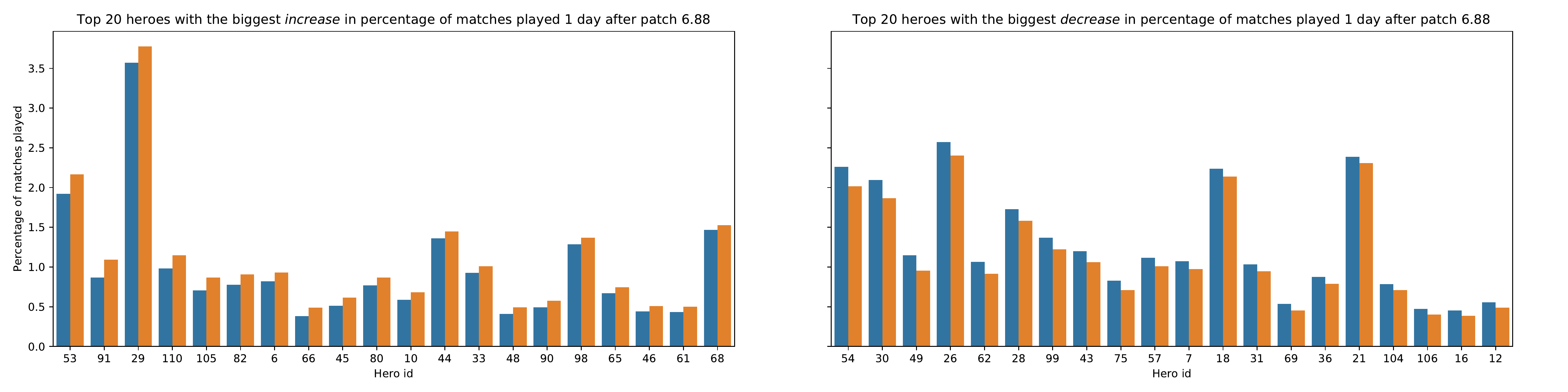}
\label{fig:changes688}
\end{figure*}

\subsubsection*{Post-patch behavior around patched characters}
Even though both patches introduced character changes, the change in character pick rates are still very different. Figure~\ref{fig:changes686} and Figure~\ref{fig:changes688} each illustrate the frequencies of character selections one day before (blue) and after (orange) patch 6.86 and 6.88. In patch 6.86, many characters saw substantial increases in the frequency of play following the patch (left pane of Figure~\ref{fig:changes686}). The increased interest in these characters among players following a patch can be attributed in part to positive changes (``buffs'') introduced by the patch. For example for the top five characters that saw the greatest increase, the most frequently-played character post-patch, Arc Warden (113), is a newly-released hero. The characters with the biggest increase in selection, Pudge (14), Zeus (22), Riki (32), and Sniper (35), can attributed buffs or reworked abilities, both of which made them stronger and more exciting to play in the new patch. Some characters saw a major decline in their frequency of play following the patch (right pane). The decreased interest in these characters among players following a patch can also be attributed in part to negative changes (``nerfs'') introduced by the patch. The five characters with the biggest decrease in selection, Windranger (21), Shadow Fiend (11), Tusk (100), Slardar (28), and Queen of Pain (39) all received explicit nerfs from the patch. Patch 6.88 brought less impactful changes on character pick frequencies. Similar trends (but different characters) are observed for characters with the biggest decrease in selection frequency in patch 6.88 as well. 

\subsubsection*{Post-patch behavior around un-patched characters}
Even though developers release patches intending to produce behavioral changes in specific characters, sometimes the consequences of those changes are difficult to predict and influence other parts of the game. Many of the characters who had significant popularity changes were not patched. Zeus (22) and Sniper (35) were un-patched characters but increased in popularity following patch 6.86. Often times, due to the synergies among the 100+ characters, changes to one character can easily lead to a cascading effect on the other character picks. What works well as counter-picks or synergy picks will not work any more. Patch 6.86 also brought other types of changes such as a reworked game map, a new type of ``rune'' that benefits certain characters greatly, reworked neutral creep abilities, some new or modified items and a new game mode. These changes all have an implicit effect on pick rates of the characters and possibly not only explain why we observed different magnitude of character pick changes in patch 6.86 and 6.88, but also why the characters that are not directly impacted also see changes in popularity. These examples demonstrate the difficulty of reliably mapping one individual or even categories of patch updates to specific behavioral changes. This uncertainty with patches in online games differentiates game patches from updates to other types of software systems where the changes may narrower and more predictable outcomes.

\subsubsection*{Post-patch behavior in online communities}
In addition to in-game behavioral changes, patches also cause disruptions to the online communities of players on social media platforms. Players in these community not only have access to the patch notes for them to reference and strategize their character picks, but also have access to online forums and discussion boards where players come together and talk about their thoughts on the new patch. For example, when patch 6.86 dropped on December 16, 2015, the announcement post on Dota2 subreddit `r/dota2' received more than 3,300 up-votes and 1,200 comments, far greater than the typical attention to other posts. Even for the comparably smaller patch 6.88, the announcement post got similar levels of attention in the community: more than 4,500 up-votes and 2,900 comments. In the comments of both announcement posts, the most popular comments include topics that highlighting notable changes, discussing potential impact the changes have and, of course, making new memes based on the changes. For example, when patch 6.86 nerfed Winter Wyvern, one of the top comments read ``Goodbye my beautiful Winter Wyvern, you were strong and fun to play, but now you have to leave us... Good night sweet princess :\'('' and the comment was followed by many other redditors agreeing and listing other characters that also received nerfs: ``WW (Winter Wyvern) and Huskar holding hands while walking to the dumpster.'' 

Besides humorous commentary, there were also serious discussions on what changes could mean to the characters. For example, in patch 6.88, a character named Pugna received a small buff on ``intelligence growth'' where one Reddit user commented: ``...the int buff combined with the change to int last patch would make his magic damage output a little higher, so maybe he'll be a bit more viable as a support?'' Sometimes, the comments are both positive and negative, like the top comment on patch 6.86 announcement post where a new hero, Arc Warden, was first added into the game: ``let the terrible reign of arc warden mis micro begin'', which both referred to the temporary popularity of a new character with players of all skill levels and the difficulties players may face when doing so, as the new character requires ``micro'', a difficult skill involving controlling multiple units simultaneously.





\subsection*{Changes in character selection behavior}\label{sec:changes-drafting-behavior}

Our first hypothesis predicted that character selection behavior should change significantly but temporarily following a patch. We measured changes in the post-patch selection behavior of players through two different constructs: the cosine similarity of players' character selections and the Gini coefficient of players' character selections.

\subsubsection*{Cosine similarity}
\begin{figure}[t]
\centering
\caption{Line plots illustrating the changes in cosine similarity (\textit{y} axis) for each of the 11 patches in the 30 days before and after the release of a patch (day 0) (\textit{x} axis).}
  \includegraphics[width=.75\textwidth]{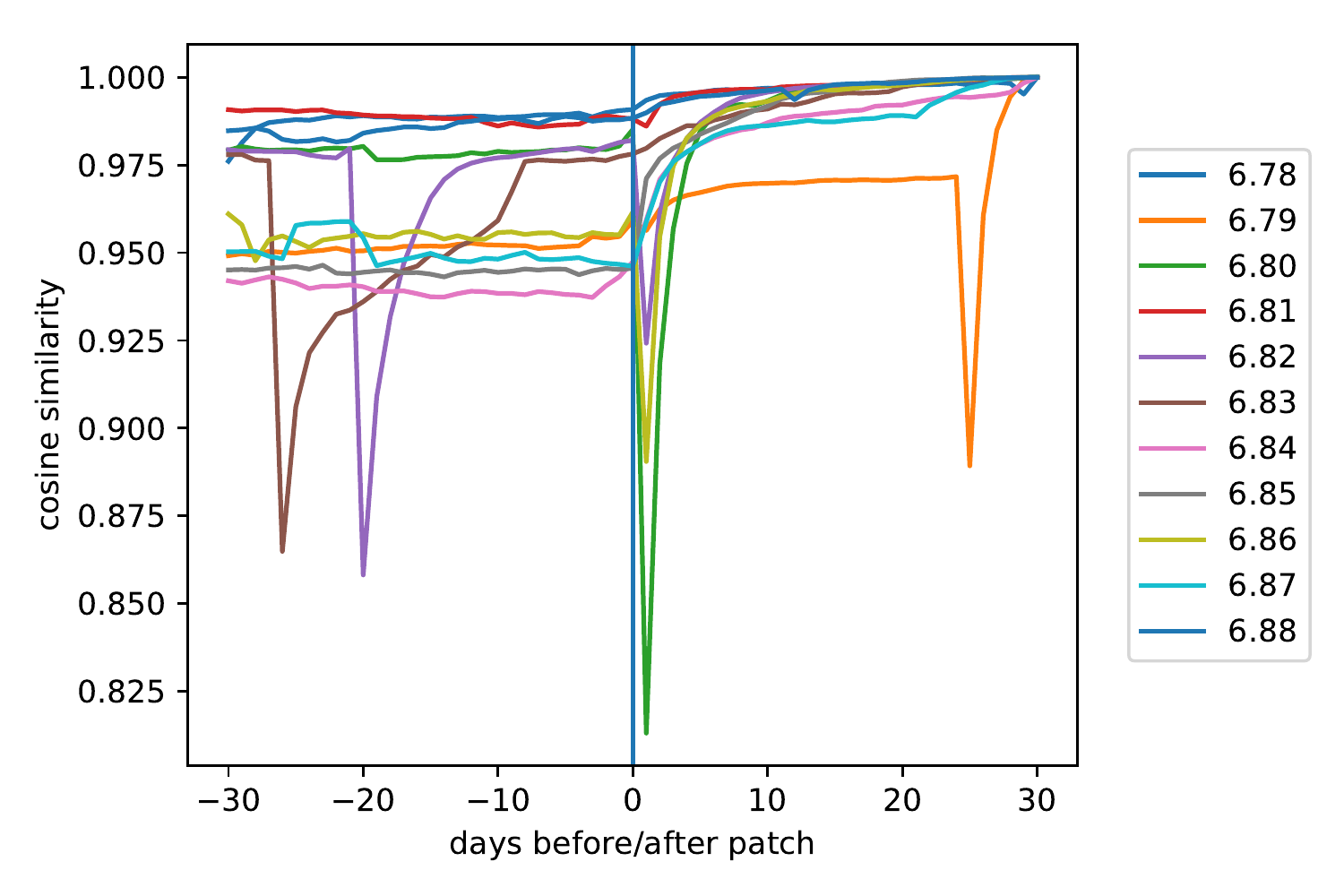}
  \label{fig:cosine}
\end{figure}

Figure~\ref{fig:cosine} shows the changes in cosine similarity of character selections for the 11 patches in the 30 days before and after a patch (patch day is time 0). Before the release of the patch, cosine similarity scores are stable for most patches except 6.82 and 6.83. This stability in cosine similarity before the patch illustrates that character selection strategies were stable within the community. One day after the release of a new patch, the cosine similarity for four patches (6.80, 6.81, 6.82 and 6.86) saw an immediate drop: $\Delta^{cosine}_{t+1} > \Delta^{cosine}_{t-1}$ (Table~\ref{tab:patch_features}). 

This sudden decrease in similarity implies that players explored new strategies by using different character selections from before the patch but few of strategies survived by the end of the 30-day post-patch window. This result can be explained by the nature of patches in \textit{Dota 2} as the changes in one category can interact with other categories. For example, when a character gets buffed directly but an item this character relies on gets nerfed, it is difficult to determine whether the net effect on the character makes them stronger or weaker after the patch. As different players explored different character selection strategies or followed influencers casting their own games, some strategies turned out to work well and players would copy this strategy. 

Tables~\ref{tab:strategies} and \ref{tab:patch_response} summarize survey responses related to team formation and character selection behavior. For the 89.5\% participants who indicated that they actively looked for information related to patches including reading patch notes (Table~\ref{tab:follow_patches}), a majority of respondents agreed (16.5\% ``strongly agree'' and 37.1\% ``somewhat agree'') with the statement ``If I am not sure a character if buffed or nerfed, I will want to try it in a game''. In addition, a super-majority of respondents agreed (24.8\% ``strongly agree'' and 53.1\% ``somewhat agree'') with the statement ``I experiment with different characters that are changed in this patch''.


There are some anomalies in Figure~\ref{fig:cosine} unrelated to patch-related behavioral changes. For example, around 25 days after the release of 6.79 (orange), the cosine similarity went down significantly, which was possibly caused by a server problem in-game where many regions ran out of servers and games could not be played. Another anomaly happened 20 days before the release of 6.82 (purple) possibly related to a major tournament, which can also led to fewer games played. We can rule out these anomalies being caused by other patches during the observation window based on the dates of patches. These anomalies follow a similar shape as patch-induced disruptions and suggest that patches are not the only mechanism that disrupt players' character selection behavior.

\subsubsection*{Gini coefficient}
\begin{figure}[t]
\centering
\caption{Line plots illustrating the changes in Gini coefficients (\textit{y} axis) for each of the 11 patches in the 30 days before and after the release of a patch (day 0) (\textit{x} axis).}
\includegraphics[width=.75\textwidth]{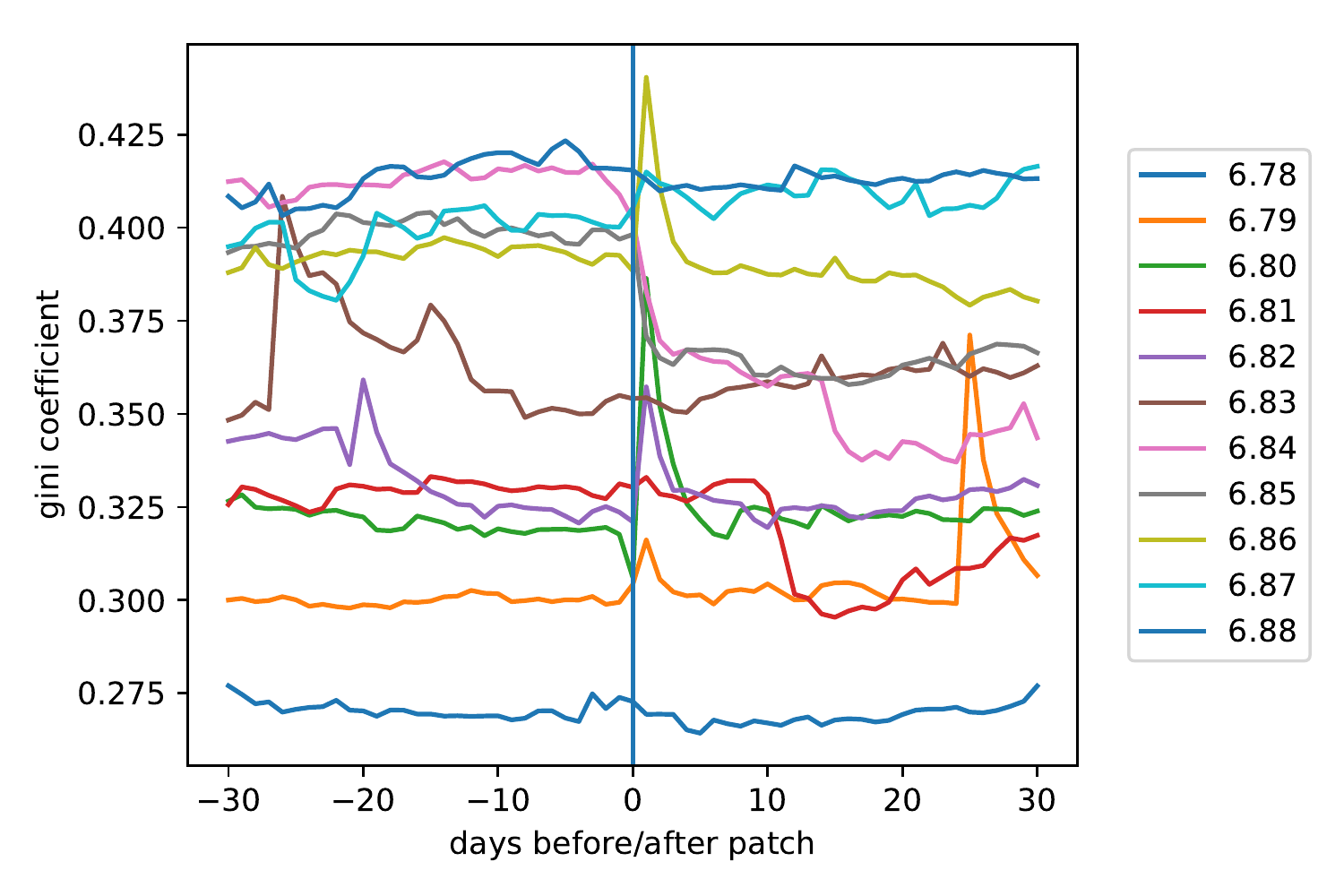}
\label{fig:gini}
\end{figure}

Figure~\ref{fig:gini} shows the changes in Gini coefficients of character selection for all 11 patches in the 30 days before and after a patch (patch day is time 0). The Gini coefficient captures whether characters are getting equal attention from players' selection behavior. Gini coefficients stayed stable for most of the patches except 6.82 and 6.83. Seven patches (6.79, 6.80, 6.81, 6.82, 6.84, 6.85, 6.86 and 6.87) saw an increase in the Gini coefficient immediately after a new patch.

Since patch notes detailing each change will be published with the release of the patches, players could be reading the changes and focusing playing the stronger characters. Out of the 98.9\% survey participants who follow patch notes, 84.2\% either strongly agree or somewhat agree that they are more likely to ban buffed characters and 87.5\% agree to some extent they are more likely to play buffed characters. In addition, 66.7\% agree they are less likely to ban a nerfed character and 64.2\% agree to some extent that they are less likely to play a nerfed character. These survey responses provide context for why players change their character selection behavior in response to patches.


If cosine similarities go down and Gini coefficients go up after a patch, it means that players are picking different characters in games and some characters are getting picked more often. If cosine similarities increase immediately after a new patch, it means that players' strategies are starting to converge to the stabilized strategies. Patch changes can be very straightforward to understand, so it requires less experimentation from players to reach their final strategies. 

\subsubsection*{Pick strategy changes}

The analysis above shows while the cosine similarities and Gini coefficients changed immediately after the patch, they do stabilize over time. To further understand how behavior changes happen and why they happen, we asked survey participants to describe their drafting strategies in two scenarios: routine drafting (``Which of the following statements best describe your hero picking strategy in general?'') versus post-patch drafting (``Which of the following statements best describe your hero picking strategy after a patch?''). Table~\ref{tab:strategies} summarizes their responses, including the changes in drafting strategies between routine and post-patch. 

\begin{table*}[t]
\footnotesize
\caption{Survey responses describing self-reported changes in team formation strategies following a patch.}
\centering
\begin{tabular}{lccc}
\toprule
\textbf{Strategy}: \textit{I pick a hero that...} &  \textbf{Routine} &  \textbf{Post-patch} &  \textbf{Difference} \\
\hline
\textit{...fills a role the team needs} &47.4\% & 36.5\% &-10.9\% \\
\textit{...I have the most experience previously playing}   &35.6\% & 26.7\% & -8.9\% \\
\textit{...is the best counter to heroes on the other team} &33.6\% & 18.9\% &-14.7\% \\
\textit{...fills a role that I have the most experience playing} &33.0\% & 28.3\% & -4.7\% \\
\textit{...I feel is the strongest in this patch} &14.0\% & 44.3\% & 30.3\% \\
\textit{...is a strong complement to another hero on the team} &9.8\% &  9.1\% & -0.7\% \\
\textit{...is most dissimilar from the other heroes on my team} & 0.4\% &  0.2\% & -0.2\% \\
\textit{Other}&10.9\% & 18.0\% &  7.1\% \\
\bottomrule
\end{tabular}
\label{tab:strategies}
\end{table*}
For the most popular character selection strategies immediately after patches (Table \ref{tab:strategies}), 44.3\% of respondents choose to pick characters that they feel are the strongest, whereas only 14.0\% responses indicated this to be their routine strategy. Another 9.4\% also pointed out that they pick whatever character is changed regardless of their changes after patches, whereas no participant indicated that this is part of their routine strategy. For routine character pick strategies, the most popular one is to prioritize filling roles on the team (47.4\%), but the percentage dropped to 36.5\% after a patch release. This shows that characters that are buffed in the patch will gain immediate attention from players, and see an increase in their pick rates. However, this phenomenon is only temporary, as indicated in Figures~\ref{fig:cosine} and~\ref{fig:gini}, because eventually players will resume their routine strategies and prioritize other drafting strategies. 

This also explains the anomalies we observed in the cosine similarities and Gini coefficients. When cosine similarity goes down and Gini coefficient goes up immediately after a new patch, it shows that players are experimenting and focusing on picking the buffed or changed characters. Since eventually players will go back to their routine pick strategies, these anomalies remain temporary. Cosine similarities start to stabilize, which means players eventually stop experimenting and start focusing on characters and strategies that work for them. Gini coefficients increase when players attend to characters that are not buffed in the new patch.

Table~\ref{tab:patch_response} summarizes survey participants' agreement with different post-patch strategies (``Think about the heroes that you play most frequently. How much do you agree with the following statements?''). The strongest agreement (strongly or somewhat agree) was found around banning buffed characters (84.2\%), drafting buffed characters (85.7\%), not banning nerfed characters (66.7\%), and not drafting nerfed characters (64.2\%). 56.7\% of respondents nevertheless agreed to some extent that they play characters regardless of buffs or nerfs introduced by a patch. The discrepancy in self-reported preferences to play buffed characters as well as playing preferred characters likely captures players' tendency to revert to a favored character if a newly-buffed character is unavailable to play because of their own team member's picks or the other team's bans. 

\begin{table*}[t]
\footnotesize
\caption{Survey responses describing self-reported changes in character selection behavior following a patch. (1--Strongly Agree; 2--Somewhat Agree; 3--Neither Agree or Disagree; 4--Somewhat Disagree; 5--Strongly Disagree)}
\centering
\begin{tabular}{l c c c c c}
\toprule
\textbf{Response} &  \textbf{1} &  \textbf{2} &  \textbf{3} &  \textbf{4} &  \textbf{5} \\
\hline
\textit{If a hero is buffed in a patch, I am more likely to play that hero.} & 37.8\% & 47.9\% &7.0\% & 4.8\% & 2.5\% \\
\textit{If a hero is nerfed in a patch, I am less likely to play that hero.} & 17.8\% & 46.4\% &   15.5\% &17.0\% & 3.3\% \\
\textit{I play heroes regardless of whether they are buffed or nerfed.}& 15.3\% & 41.4\% &   21.1\% &15.8\% & 6.5\% \\
\textit{I am more likely to ban a buffed hero.}  & 52.9\% & 31.3\% &6.5\% & 6.0\% & 3.3\% \\
\textit{I am less likely to ban a nerfed hero.}  & 34.1\% & 32.6\% & 14.5\% &12.0\% & 6.8\% \\
\textit{I ban heroes regardless of their changes}  & 17.8\% & 30.8\% & 21.8\% &18.8\% & 10.8\% \\
\textit{I change hero picks based on the item changes introduced by a patch.} & 12.0\% & 44.6\% & 19.5\% &18.3\% & 6.5\% \\
\bottomrule
\end{tabular}
\label{tab:patch_response}
\end{table*}


\subsection*{Relationship between patch severity and behavioral changes}

\begin{figure*}[t]
    \centering
    \caption{Regression plots illustrating the relationship between patch severity (\textit{x} axes) and the four types of behavioral changes (\textit{y} axes). The shaded area is the 95\% confidence interval. Regression coefficients ($\beta$), p-values ($p$), and r-squared values ($r$) for each relationship are reported in the lower right corners.}
    \includegraphics[width=\textwidth]{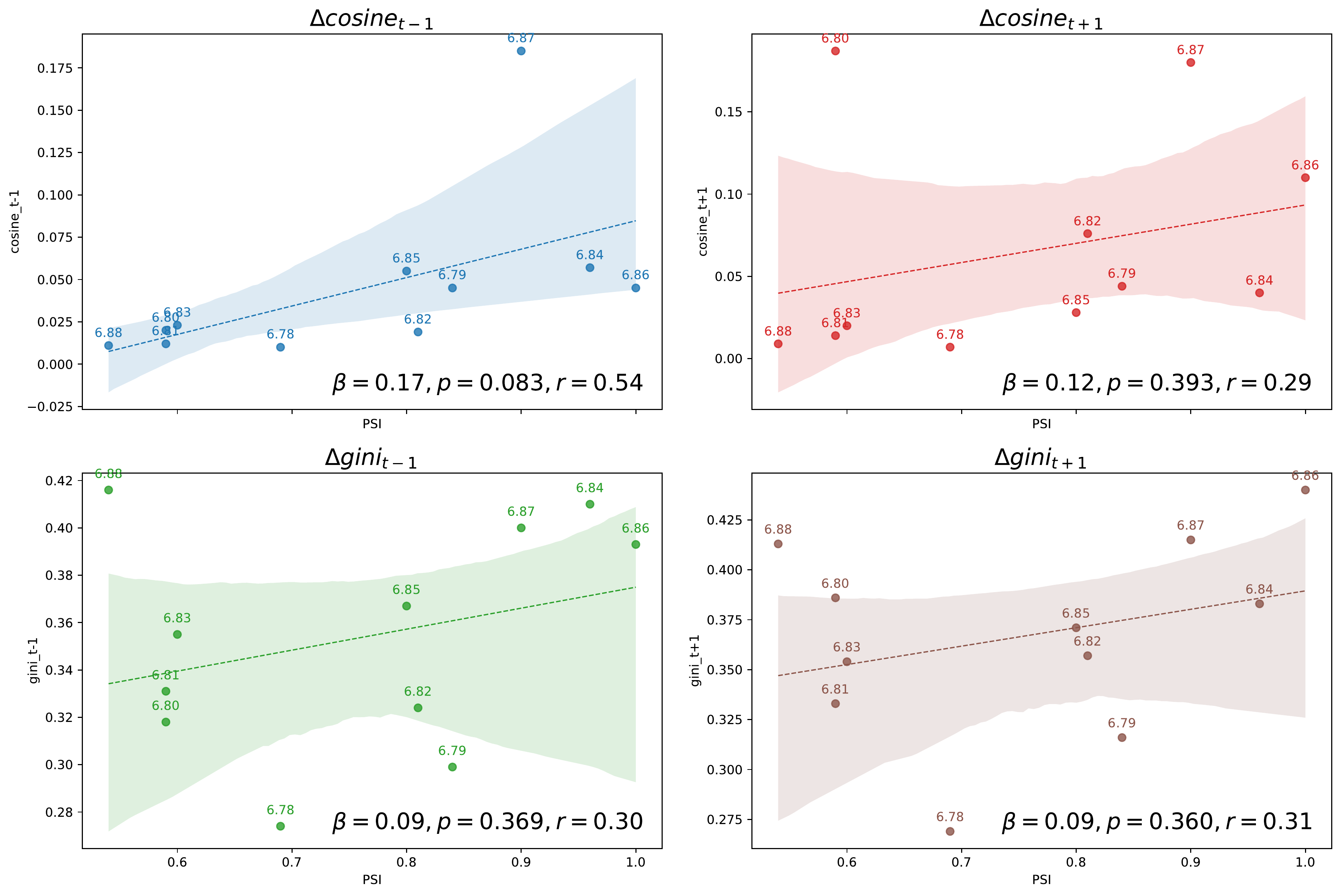}
    \label{fig:psi_behavior_changes}
\end{figure*}

Our second hypothesis predicted more severe patches will have larger changes in drafting behavior. Table~\ref{tab:patch_features} summarizes the aggregate changes in player behaviors ($\Delta{cosine}$ and $gini$) against the changes introduced by the eleven patches in our sample (PSI and Total Changes) using the features we extracted from each patch time series described in Figure~\ref{fig:delta_example}. Figure~\ref{fig:psi_behavior_changes} shows the changes in cosine similarity (top, blue and red) and Gini coefficient (bottom, green and brown) in response to patches. A best-fit regression line (dotted lines) with 95\% confidence intervals (shaded areas) are estimated for each relationship. The linear regression coefficient ($\beta$), significance ($p$), and explained variance ($r$) are reported in the lower right corner of each pane. We observe positive coefficients and moderate r-values for all four behavior changes, though no relationship is statistically significant at a $p<0.05$ level because the small number of observations makes the linear regression models especially sensitive to outliers.

$\Delta_{t-1}^{cosine}$ is the change in cosine similarity from the day preceding the patch to 30 days after the patch: this captures the change in the similarity of character selection behavior from \textit{before} the patch. The positive relationship ($\beta=0.17, p=0.083, r=0.54$) indicates the more severe a patch is, the more dissimilar character selection behaviors are compared to before the patch. Patch 6.87 is a major outlier from the trend with major changes to the drafting phase, major changes to the non-player characters and buildings, and base statistics. The weights we derived from survey respondents assigned low importance to these types of changes, which could be a reason why the calculated patch severity does not scale with the observed behavioral change.

$\Delta_{t+1}^{cosine}$ is the change in cosine similarity from the day following the patch to 30 days after the patch: this captures the change in behavior immediately \textit{after} the patch. The positive relationship ($\beta=0.12, p=0.393, r=0.29$) indicates the more severe a patch is, the more dissimilar character selection behaviors are compared to after the patch. Both 6.80 and 6.87 saw major changes to the timing (6.80) and picking/banning voting process (6.87) in the character selection phase, which could explain their anomalously large post-patch changes relative to other patches. 

$\Delta_{t-1}^{gini}$ is the change in Gini coefficient from the day preceding the patch to 30 days after the patch: this captures the change in the concentration of character selection behavior from \textit{before} the patch. The positive relationship ($\beta=0.09, p=0.369, r=0.30$) indicates the more severe a patch is, the more concentrated character selection behaviors became compared to before the patch. $\Delta_{t+1}^{gini}$ is the change in Gini coefficient from the day following the patch to 30 days after the patch: this captures the change in the concentration of character selection behavior from \textit{after} the patch. The positive relationship ($\beta=0.09, p=0.360, r=0.31$) indicates the more severe a patch is, the more concentrated character selection behaviors became compared to after the patch. 

Patches 6.86, 6.87, and 6.88 had anomalously high changes in Gini coefficients relative to their calculated patch severity scores. Patch 6.87 continues to be an outlier for all behavior change measures, for reasons related to those previously discussed above. Patch 6.88 only had changes in the ``Hero'' and ``Items'' categories with the 0 changes in other categories likely over-deflating its computed severity score. Patch 6.86 saw major reworks to six heroes and their temporary removal from the ``Captain's Mode'' game mode, one of four modes we included in our analysis, that may have shifted the concentration of character selections following the patch. Patches 6.78 and 6.79 had anomalously low changes in Gini coefficients relative to their calculated patch severity scores. 6.78 had significant changes to the selection phase of Captain's Mode and 6.79 had many changes in the Neutral category. These changes may have over-inflated the calculated severity score relative to the observed behavioral change.

\section*{Discussion}
Software patches are an overlooked driver of behavioral change on social platforms. Because patches introduce changes to the rules and incentives within multiplayer online games, players of these games adapt their tactics and learn new strategies in response. We examined the corpus of detailed patch notes of the popular MOBA game \textit{Dota 2} and compared the number and types of changes introduced by a patch with the changes in players' in-game behavior--character selections. We hypothesized that drafting behavior would change significantly and temporarily following a patch (H1) and more severe patches would have larger changes in drafting behavior (H2). We employed a mixture of methods to understand the role of software patches on player behavior, including a content analysis of 11 major patch notes between 2013 and 2016, log data analysis of character selection decisions in 532 million matches in the 30 days before and after each of these patches, and a survey of 437 players to contextualize the patterns from the log data. Our analysis found evidence of substantial behavioral changes following the release of a new patch and the magnitude of these behavioral changes are scale with the calculated ``severity'' of the patch. 

Disruptions like software patches are compelling empirical settings to study the relationships between technology use and social behavior. While software patches are released with the intent to change user behavior by changing incentives, removing imbalances, and introducing new features, these changes occur within an extremely complex and multi-dimensional decision space. Changes to any character, item, or game mechanic can have unintended consequences that are discovered, exploited, and can ruin the experience for hundreds of thousands of players around the world. Patches are important para-texts~\citep{carter_emitexts_2015,consalvo_paratext_2017} that not only structure the experience of being a player who attends to patch changes to remain competitive but also being part of a community where patch changes are occasions to converge, discuss, learn, and collaborate. While our findings primarily emphasize the behavioral changes occurring within the game, the case study highlights the coupling of this in-game behavior with content production and engagement in online communities. As social computing research shifts to ecological perspectives about how users and content move across multiple social platforms~\citep{zhao_ecology_2016,hall_crossplatform_2018,wilson2020cross,zuckerman_ecosystems_2021}, software patches in online games are promising empirical settings with well-identified disruptions for studying cross-platform behaviors.

Because patches introduce dozens or hundreds of interacting changes simultaneously into this multi-dimensional space, any of which cause significant changes in behavior, this makes precise causal identification difficult~\citep{hill_measuring_2015}. This study employed a simple research design focused on the aggregated behavioral changes following software patches. The combination of multiple sources of change, the precise documentation of what changed, and the detailed and unobtrusively-gathered longitudinal observations involving thousands of users makes ``patch-induced behavioral change'' in online games like MOBAs extremely rich sites for theory-testing and methodological development. More complex analyses stratifying players by skill or geography, incorporating other disruptions like tournaments, or classifying patch changes into more nuanced categories are worthy topics of future research. Our analyses combining survey and log data only begin to scratch the surface of this potential but illustrate how players weigh changes to characters more than other patch changes. These findings are important guides for researchers to orient to the kinds of changes and behaviors to prioritize given the multiple sources of variation, complexity of behavior, and scale of data.

The responses of gaming communities to software patches are also unique opportunities to observe the confluence of cultural norms, social behaviors, and technical practices for maintaining complex socio-technical infrastructures~\citep{ribes2010sociotechnical,lee_cyberinfrastructure_2006,plantin2018infrastructure}. Unlike software changes in office productivity or mobile apps that occur in the background, patches in MOBAs and other online games are significant disruptions that are eagerly anticipated, hotly debated, and collaboratively indexed~\citep{bainbridge_creative_2007,gursoy_understanding_2019}. The importance of software patches within gaming communities illuminates the socio-material relationships of these seemingly banal technical artifacts as not only sites of infrastructural power and maintenance~\citep{chapin_types_2001,vaniea_tales_2016,bergman_cognitive_2018} but also occasions for organizing and even resistance~\citep{acker_software_2016,svelch_resisting_2019,morreale_update_2020}. Our findings about patch-induced behavioral changes contribute to this emerging strand of research and invite us to revisit of prevailing conceptions of users of passive consumers of software engineering practices and instead see them as actively influencing software engineering practices. 

\subsection*{Implications}
The patch severity index (PSI) can serve as a new tool to analyze the impacts of software patches. The PSI can be extended methodologically by using alternative data collection and weighting strategies. While the PSI might be adapted to other online games and social-technical systems to understand the consequences of changes on player behavior, software patches in other sociotechnical systems do not necessarily induce learning or mimicry among their users. Patches are not a one-directional relationship: developers react to analytics about player engagement or character win-rates and players actively lobby for developers to implement changes. These feedback loops between developers and user communities remain poorly understood and even less theorized. A well-validated PSI could be used by developers to identify potential sources of community resistance or acceptance to software changes. Efforts to utilize metrics like a PSI could lead to more ``human-centered'' and social software engineering strategies that invites greater attention to how users and communities interpret and adopt the changes introduced by patches. 

Multiplayer online battle arenas and esports generally also have enormous cultural and economic influence~\citep{Wingfield2014Aug,reitman_esports_2020,seo_professionalized_2016}. Understanding the contexts and consequences of software patches on gameplay is thus of increasing consequence for elite players, their managers and owners, audiences, as well as game developers and the general player community. Developing a systematic understanding of the effects of software patches could improve the ability for professional organizations to identify new strategies and train more effectively.

Game developers deploy patches to introduce new features, correct errors, and alter incentives to play the game. Because patches change the fundamental rules and structures of the game, they destabilize players' deep investment in developing heuristics, tactics, and strategies about how to play the game. This coordinated destabilization of meaning has consequences for hundreds of thousands to millions of people who collectively engage in sensemaking to understand what has changed, exploration of new possibilities within an enormous and complex decision space, and learning new strategies and possibilities~\citep{weick2005organizing}. Patches thus provide a rich empirical setting to explore processes of naturalistic decision-making~\citep{gore_naturalistic_2006}, temporary and high-tempo team processes~\citep{bakker_taking_2010,bechky_gaffers_2006,faraj_coordination_2006}, and organizational learning~\citep{march_exploration_1991,lazer_exploration_2007} unfolding on a cross-cultural scale involving hundreds of thousands or millions of people.

\subsection*{Limitations and future work}
There are some limitations to our analysis. First, we make no claims about the generalizability of these results to other games or software contexts: the results should only be interpreted in the context of this large-scale sample of 532 million matches and survey responses for \textit{Dota 2} within this specific time frame. Moreover, this implementation of the PSI may not be generalizable to other genres of games. However, the PSI as a method can inspire ways to quantify the impact of software patches by leveraging patch documentation and community feedback on previous patches. Although measuring post-patch changes using software log data will also show the overall impact of patches, it is difficult to notice the correlations between the types of changes and the potential magnitude of impact for each type of changes.  By categorizing potential changes and surveying community reactions towards each category, developers can make more informed decisions about what to prioritize, how to document changes, and what behavioral changes to monitor. 

Second, patches in \textit{Dota 2} often introduce many changes in different categories at the same time, and it is very difficult to causally isolate the relationship between the types of changes and the magnitude of behavior changes. A newly-introduced character may be overpowered, a popular character may be nerfed, and underutilized items may be buffed. These changes all interact with each other in the form of drafts, bans, and in-match tactics that are simultaneously explored and exploited by players. While we have analyzed over 532 million matches in this analysis, our primary unit of analysis were the 11 major patches. Including changes in character selection behavior for patches after 2016 could provide additional statistical power for analyses like the relationship between patch severity and behavioral changes. Alternative implementations of the Patch Severity Index, for example emphasizing character changes or determining weights on other criteria, might recover stronger relationships between patches and changes in player behavior. 

Our analysis of player behavior aggregated matches because the data lacked sufficient detail and coverage to stratify for player skill, history, identity, or geography, all of which can bias estimates through misaggregation~\citep{barbosa_averaging_2016}. Prior research has explored the significant differences in team assembly strategies across skill levels~\citep{kim_proficiency_2016} and effects of play session length~\citep{sapienza_dynamics_2017}. Future work should look at specific game modes, such as Captain's Mode, where there is a standardized pick-and-ban phase and players' team composition strategies have more impact on their drafts. We would expect the changes in character picks to be more stark, even for patches that did not show significant aggregate changes. We  We particularly emphasize the importance of comparing draft changes across players' skill brackets as post-patch behavioral changes are likely to vary strongly based on skill and tactics.

Finally, this analysis did not explore the role that online communities and social content play in behavioral changes. Players use social platforms like discussion forums, news aggregators, wikis, and streaming sites to produce, revise, and consume information about patches. The case study suggests software patches not only disrupt in-game player behavior but also social content production in these communities. Understanding the coupling between the volume and content of social content production like patch analysis videos could illuminate richer cross-platform dynamics about patch-induced behavioral change. Furthermore, tournaments, championships, and streams from elite players are alternative sources of influence where innovative new tactics, builds, and draft strategies could drive significant changes in aggregate player behavior. Future research studying player behavior in online games should also consider software patches' impact for players in-game and in the online gaming communities.

\section*{Conclusion}
Game developers' releases of software patches and the behavioral effects of patches on millions of players is a promising empirical setting for understanding more general processes like the social consequences of technological disruptions. Unlike most other user communities, software patches are a prominent and important part of the gaming culture around multiplayer online battle arena games like \textit{Dota 2}. Using a mixture of quantitative methods including a survey, text analysis, and log data analysis, we characterized how players' character selection behavior changed following patches, developed a method for quantifying patch severity, and the identified a significant relationship between patch severity and post-patch behavioral disruption. We encourage future research to identify the mechanisms linking ``patch severity'' and behavioral disruptions in other settings and how players' responses to patch-induced disruptions can be models for improving the resilience of social systems.

\section*{Acknowledgments}


\bibliographystyle{ACM-Reference-Format}
\bibliography{bibliography}


\end{document}